\documentclass[conference]{IEEEtran}
\IEEEoverridecommandlockouts
\usepackage{cite}
\usepackage{fancyhdr}
\usepackage{amsmath,amssymb,amsfonts}
\usepackage{algorithmic}
\usepackage{graphicx}
\usepackage{textcomp}
\usepackage[dvipsnames, svgnames]{xcolor}
\usepackage{listings}
\usepackage[most]{tcolorbox}

\usepackage{etoolbox}
\usepackage{tikz} %

\makeatletter
\patchcmd{\@maketitle}
  {\addvspace{0.5\baselineskip}\egroup}
  {\addvspace{-1.0\baselineskip}\egroup} %
  {}
  {}
\makeatother

\definecolor{codeGreen}{rgb}{0.0,0.5,0.0}
\definecolor{codeBlue}{rgb}{0.0,0.2,0.8}
\definecolor{codeGray}{rgb}{0.4,0.4,0.4}

\newtcblisting{IEEEcodeboxCompact}[2][]{
  enhanced,
  colback=gray!5,
  colframe=gray!60!black,
  colbacktitle=gray!20!white,
  coltitle=black,
  fonttitle=\bfseries\sffamily\scriptsize,
  title={#2},
  arc=1pt,
  boxrule=0.6pt,
  left=3mm, right=1mm, top=0mm, bottom=0mm, %
  listing only,
  listing options={
    language=C++,
    basicstyle=\ttfamily\scriptsize,          %
    keywordstyle=\color{codeBlue}\bfseries,
    commentstyle=\color{codeGreen}\itshape,
    numbers=left,
    numberstyle=\tiny\color{codeGray},
    numbersep=5pt,
    tabsize=2,
    aboveskip=0pt, %
    belowskip=0pt, %
    morekeywords={MPI_Comm, size_t}
  },
  #1
}
\usepackage{cellspace}
\ifx \debug \undefined
  \newcommand{\roger}[1]{{}}
  \newcommand{\ej}[1]{{}}
  \newcommand{\noel}[1]{{}}
  \newcommand{\trevor}[1]{{}}
  \newcommand{\lance}[1]{{}}
\else
  \newcommand{\roger}[1]{{\textcolor{DarkOrchid}{Roger: #1}}}
  \newcommand{\ej}[1]{{\textcolor{DeepPink}{EJ: #1}}}
  \newcommand{\noel}[1]{{\textcolor{DarkOrange}{Noel: #1}}}
  \newcommand{\trevor}[1]{{\textcolor{LimeGreen}{Trevor: #1}}}
  \newcommand{\lance}[1]{{\textcolor{ForestGreen}{Lance: #1}}}
\fi

\begin{document}

\title{MailRail: Mitigating Incast Congestion with User-Space, Node-Based Flow Control}

\author{
\IEEEauthorblockN{Noel Gomillion\textsuperscript{\dag\ddag}, 
                  Lance Fletcher\textsuperscript{\ddag}, 
                  Trevor Steil\textsuperscript{\dag}, 
                  Eun Jung Kim\textsuperscript{\ddag}, 
                  Roger Pearce\textsuperscript{\dag\ddag}}
\IEEEauthorblockA{\textsuperscript{\dag}\textit{Center for Applied Scientific Computing}; 
\textit{Lawrence Livermore National Laboratory}}
\IEEEauthorblockA{\textsuperscript{\ddag}\textit{Department of Computer Science and Engineering}; 
\textit{Texas A\&M University}\\
\{gomillion1, steil1, rpearce\}@llnl.gov, \{ngomillion, lance.fletcher, ejkim, rpearce\}@tamu.edu}
}
    \maketitle
\thispagestyle{fancy} 
\lhead{} 
\rhead{} 
\chead{} 
\lfoot{\footnotesize{SC26 Workshops, November 15-20, 2026, Chicago, Illinois, USA 
\newline 979-8-3195-1221-5/26/\$31.00 \copyright 2026 IEEE}} 
\rfoot{} 
\cfoot{} 
\renewcommand{\headrulewidth}{0pt} 
\renewcommand{\footrulewidth}{0pt} 

\begin{abstract}
Irregular workloads, such as those stemming from Graph Neural Networks (GNNs) and Mixture-of-Experts (MoE) models, frequently induce severe incast congestion in large-scale parallel environments. Existing fabric-level congestion control mechanisms rely on reactive, link-level rate limiting that degrades overall throughput. While receiver-driven, credit-based flow control offers a proactive alternative, as process counts scale memory slots per rank fall below usable thresholds, rendering traditional rank-based accounting untenable.
To address this challenge, we introduce \textbf{MailRail}, an out-of-order, asynchronous messaging framework, along with a novel \textbf{user-space, node-based credit flow control} system. By shifting credit accounting from individual processes to shared per-node pools, our proactive, fabric-agnostic approach maintains sufficient credit availability even as process density grows, enabling processes to dynamically utilize freed credits without waiting for receiver-side redistribution. Experimental evaluations demonstrate that node-based flow control achieves up to a \textbf{$1132\times$ speedup} over unconstrained execution at an adversarial incast ratio of 4096:1 while adding minimal overhead on uniform workloads.
\end{abstract}

\section{Introduction}
Both Graph Neural Networks (GNNs) and Mixture-of-Experts (MoE) models generate highly irregular communication patterns that are inherently prone to inducing incast congestion, a severe bottleneck where many sender processes simultaneously flood a small set of receiving victim processes \cite{MoE,GNN,Graph}. GNNs use real-world datasets for training, which typically exhibit power-law degree distributions \cite{power-law,real-world}. During training, aggregating features from hub nodes causes fan-in traffic unless mitigation techniques are used. In MoE models, used to make AI inference more scalable, tokens are routed to a subset of experts. Some experts prove more popular than others, creating many-to-one traffic spikes at GPUs hosting those experts during All-to-All collectives \cite{MoE}. 

As one of the most destructive forms of network congestion in distributed architectures, incast congestion causes severe communication stalls, leaving high-speed compute and memory resources underutilized and significantly degrading overall system throughput to near zero \cite{rdfc,Incast}. In lossless network fabrics, link-level flow control prevents packet loss by backpressuring senders when receiver buffers saturate. During incast communication, the victims’ buffers quickly saturate, forcing upstream senders to stall. Congestion spreading ensues, causing severe performance degradation. Thus, congestion control represents a critical challenge in high-performance computing (HPC). Current incast mitigation strategies focus on hardware-level solutions that are reactive instead of proactive. These include sender rate limiting, which has been shown to degrade effective throughput rates \cite{MoE,rdfc}. 

To overcome these limitations, end-to-end, receiver-driven, credit-based flow control offers a proactive alternative by tying transmission credits directly to pre-allocated buffer memory at the receiver, bounding in-flight traffic before congestion occurs \cite{rdcbfc, protego}. However, as supercomputer core counts and process densities scale, traditional rank-based credit accounting faces severe memory constraints: dividing receiver memory among thousands of sending processes reduces available credits to less than one buffer slot per rank, rendering rank-based accounting untenable. 

While traditional rank-based flow control manages credits on a per-process (MPI rank) basis, node-based flow control aggregates credit accounting into a shared, per-node pool accessed dynamically by all local ranks on the same physical host. By shifting credit accounting from individual processes to shared per-node pools, local sending ranks can dynamically access a collective quota of credits, maintaining high throughput and credit availability even when receiver memory slots fall below one per rank.
Although receiver-side credit redistribution strategies attempt to adapt to shifting process demands \cite{rdfc, UEC, dynamic,gasnet}, they remain inherently reactive by relying on the receiver to detect communication bottlenecks and adjust allocations across the network. In contrast, sender-side node-based credit pooling naturally reallocates available credits instantaneously as local processes complete their transfers.

In this work, we introduce MailRail, an asynchronous messaging framework designed for high-performance out-of-order communication. Within MailRail, we propose two end-to-end, user-space, credit-based flow control mechanisms to address incast congestion in a fabric-agnostic manner: rank-based flow control, which manages credits per process, and node-based flow control, which aggregates credits into host-level shared pools.
We evaluate MailRail across synthetic benchmarks and real-world application traces, including an adversarial incast benchmark, a uniform-traffic histogram workload, and trace replays from YGM, LLNL’s asynchronous distributed active message runtime  \cite{YGM}. Experimental results show that node-based flow control consistently outperforms traditional rank-based and unconstrained execution, achieving up to a $1132\times$ speedup at an adversarial incast ratio of 4096:1, sustaining consistent goodput under extreme 4096:1 incast ratios, and demonstrating linear runtime scaling as the workload per victim increases.

The main contributions of this work are:
\begin{itemize}
    
    \item an asynchronous messaging API for out-of-order communication; 
    \item two end-to-end, user-space, receiver-driven, credit-based flow control solutions to the incast communication problem that are network-fabric agnostic: rank-based flow control and node-based flow control;
    \item an adversarial benchmark for testing incast communication performance under various incast ratios;
    \item an analysis of incast sensitivity during unconstrained execution on the supercomputers Dane and Lynx.
\end{itemize}

The remainder of this paper is organized as follows: Section~\ref{sec:background} discusses background and related work on congestion control mechanisms. Section~\ref{sec:design} presents the MailRail messaging interface, while Section~\ref{sec:implementation} details our rank-based and node-based flow control implementations.
Section~\ref{sec:evaluation} describes our experimental methodology and performance evaluations across synthetic benchmarks and application traces. Finally, Section~\ref{sec:conclusion} concludes the paper and outlines future work.

\section{Background}
\label{sec:background}

Incast communication, where many processes send to a few victims, rapidly saturates receiver buffer space at victim processes, causing severe network backpressure, buffer bloat, and throughput collapse that hinders system scalability. Prior mitigation approaches predominantly rely on hardware-level or sender-side rate-limiting mechanisms to manage these bottlenecks, including the following:
\subsection{Network Fabric Strategies}
\subsubsection{Ethernet with RoCEv2}
RoCEv2 enforces losslessness using Priority-based Flow Control (PFC) at the link level and regulates send rate via Data Center Quantized Congestion Notification (DCQCN). When incast communication saturates switch buffers, DCQCN relies on feedback loops using Explicit Congestion Notification (ECN) and Congestion Notification Packets (CNP) to throttle senders. These feedback loops take multiple round-trip times (RTT) to take effect. This delayed reaction causes throughput oscillations, buffer overruns, PFC pause storms, and Head-of-Line (HoL) blocking across the fabric \cite{cornelis_whitepaper,broadcom,datacenter}.

\subsubsection{Receiver-Driven Congestion Control for InfiniBand}
To prevent packet loss, InfiniBand (IB) utilizes hop-by-hop credit-based flow control. However, fan-in incast rapidly depletes switch credits, causing backpressure to cascade upstream and freeze even uncongested flows. Receiver-driven paradigms resolve this by granting transmission windows and credit tokens from the receiving host \cite{rdfc}.

\subsubsection{Ultra Ethernet Consortium}
The Ultra Ethernet Transport (UET) implements Receiver Credit-Based Congestion Control (RCCC) to eliminate last-hop incast. The receiving host tracks active senders and allocates bandwidth using microsecond time slots via ACK\_CC control packets \cite{UEC}.

\subsubsection{Omni-Path Express (OPX)}

Omni-Path Express (OPX) mitigates receiver incast using in-fabric telemetry and flit-level credits rather than the link pauses used by PFC. When switch buffers near the receiving host fill up, switches issue backward notifications to throttle the sending hosts, causing the congestion for that target endpoint; the sender is free to continue sending to uncongested recipients because only the traffic sent to the congested endpoint is throttled. Per-virtual-lane credit accounting caps ingress data volume to prevent buffer overruns \cite{cornelis_whitepaper}.

\subsection{Endpoint Software Strategies}

\subsubsection{MPI Protocols}
Message Passing Interface (MPI) is a library specification widely used for communication in HPC systems. MPI utilizes protocols that require acknowledgment handshakes to limit sender data transmission. This includes the Rendezvous protocol, which is triggered when a message exceeds a given size, and synchronous sends (e.g. MPI\_Ssend and MPI\_Issend), which wait for a receipt acknowledgment before the communication can finish. Both of these techniques introduce additional RTT latency, though MPI\_Issend can overlap this delay with computation \cite{MPI, mpi41}.
\subsubsection{Flow Control for GASNet-EX}
GASNet-EX is a communication library for the network middleware layer designed for Active Messaging (AM) and Remote Memory Access (RMA). As part of the library, GASNet-EX uses flow control that ties the number of credits each sending process has to available buffer space on the receiving process using a dynamic credit allocation scheme that initially underallocates credits, then reallocates surplus credits to processes that have previously stalled due to credit starvation when sending AM Requests \cite{gasnet}.

\subsection{Motivation}

While irregular workloads achieve higher throughput using asynchronous execution, small asynchronous messages can cause congestion, buffer overflows, and out-of-memory (OOM) errors. Examined congestion control strategies fail to prevent this without negatively impacting effective throughput: sender rate limiting experiences delayed responses and prolonged recovery times, while software protocols requiring handshakes or dynamic receiver-side adjustments create RTT delays. Our proposed node-based flow control overcomes these limits by proactively shaping traffic flow at the user-space level while naturally allocating credits to active sending processes. This enables consistent throughput without triggering switch backpressure regardless of the underlying network fabric.

\section{MailRail Interface}
\label{sec:design}

\begin{figure}[t]
\centering
\begin{IEEEcodeboxCompact}{MailRail C++ Class Interface}
class mailrail {
public:
  // Lifecycle
  int mailrail_init(MPI_Comm comm);
  int mailrail_finalize();

  // Zero-Copy Send Path (Backpressure)
  int mailrail_reserve_send_buffer(int dest, void **ptr,
                              int *capacity);
  int mailrail_send(void *ptr, int size);

  // Zero-Copy Receive Path & Progress
  int mailrail_receive(void **ptr, int *size);
  int mailrail_release_receive_buffer(void *ptr);
};
\end{IEEEcodeboxCompact}
\caption{The MailRail API allows for zero-copy sends and receives while backpressuring the sender when buffer space is unavailable. While the MailRail API is conceptually transport-agnostic, our current implementation uses MPI as the underlying communication layer. All MPI-specific dependencies are limited to the init call; the remaining data-path functions operate on untyped memory buffers, allowing the underlying transport to be swapped out in future.}
\label{fig:code_api}
\end{figure}

The MailRail API is simple by design (see Fig.~\ref{fig:code_api}), allowing for experiments targeting fire-and-forget asynchronous communication \cite{YGM}. This communication model exploits one-sided, non-blocking asynchronous messages to eliminate round-trip latency associated with handshakes and acknowledgments. Message processing order reflects arrival order at the receiver, independent of send order or sender identity. MailRail does not provide a mechanism for synchronization and requires user code to handle quiescence detection. The underlying messaging layer is abstracted, enabling implementations for various communication substrates (e.g., MPI, UCX, or libfabric).

To deliver high throughput without incurring memory copy overhead, the MailRail API centers on these core principles: zero-copy, sender-side backpressure, and user-driven progress.

\subsection{Buffer Reservation and Sender-side Backpressure}

To achieve zero-copy performance, MailRail requires applications to request a send buffer via \text{mailrail\_reserve\_send\_buffer} prior to populating payload data. Crucially, buffer allocation failure is not an error condition; rather, it serves as MailRail's primary sender-side backpressure mechanism (see Fig.~\ref{fig:sending_code}). When internal buffers are saturated, the call returns a null buffer and immediately yields execution back to the caller. Applications handle saturation through cooperative polling, interleaving reservation attempts with calls to \text{mailrail\_receive}. This ensures a sender stalled by backpressure continues to drain its incoming queues, avoiding distributed deadlocks.

Once populated, the message is relayed via \text{mailrail\_send}. This operation is nonblocking and fire-and-forget. Depending on the implementation, a message is not necessarily sent immediately when this function is called. The API allows for sends to be deferred in a deferred send queue. When implemented, this would enable the application to continue execution without waiting for a send to complete, which is in keeping with our asynchronous model. 

\begin{figure}[t]
\centering
\begin{IEEEcodeboxCompact}{Example of Sending Messages}
mailrail mb;
int result = mailrail_init(MPI_COMM_WORLD);
void* send_buffer = nullptr;
int capacity = datagram_size;
while( result != 0 || send_buffer == NULL ) {
    result = mb.mailrail_reserve_send_buffer(dest, 
                            &send_buffer, &capacity);  
    receive_message_poll();
}
fill_buffer(send_buffer);
mb.mailrail_send(send_buffer, num_bytes);
\end{IEEEcodeboxCompact}
\caption{This example shows how to reserve and send buffers. By interleaving reservation attempts with calls to \text{mailrail\_receive}, senders continue to drain incoming queues while waiting for an available buffer.}
\label{fig:sending_code}
\end{figure}

\subsection{Zero-copy Receive and User-driven Progress}

Messages are received using nonblocking calls to \text{mailrail\_receive}. When a message is present, MailRail returns a direct pointer into its own internal receive pool, along with the payload size. Calls to \text{mailrail\_receive} double as the runtime's internal progress engine, draining deferred outbound sends while processing incoming messages (see Fig.~\ref{fig:polling_code}). After the application finishes processing the buffer, the application signals MailRail to return the buffer to the receive buffer pool via \text{mailrail\_release\_receive\_buffer}, which also posts new non-blocking receive requests. MailRail uses \text{mailrail\_finalize}, a collective operation performed by all ranks together, to cancel any pending communication and to release buffers.

\begin{figure}[t]
\centering
\begin{IEEEcodeboxCompact}{Example of Polling for Received Messages}
void* recv_ptr = NULL;
int recv_size = 0;
do{
  result = mb.mailrail_receive(&recv_ptr, &recv_size);
  if (recv_ptr != NULL && recv_size > 0) {
    received_message = 1;
    process_received_message(recv_ptr, recv_size);
    result=mb.mailrail_release_receive_buffer(recv_ptr);
  } else {
    received_message = 0;
  }
} while(received_message);
\end{IEEEcodeboxCompact}
\caption{Aggressively receiving messages acts as the user-driven progress mechanism; this clears incoming receive queues as quickly as possible while also draining any pending send messages MailRail may have.}
  \label{fig:polling_code}
\end{figure}

\section{MailRail Implementation}
\label{sec:implementation}

MailRail's simplicity allows for various implementation strategies and design decisions. This enables experimentation over a wide variety of communication paradigms. Here we discuss details of a MailRail implementation centered on MPI, including node- and rank-based flow control. We first discuss general implementation decisions, then the specifics of each flow control mechanism.

\subsection{MailRail over MPI}

 We implement MailRail using MPI. Because we focus on asynchronous communication, we use the nonblocking functions \text{MPI\_Isend} and \text{MPI\_Irecv} for passing messages. Receivers process messages in the order they arrive without regard to the source or the message tag.
 
To avoid malloc overhead, MailRail reuses memory through dedicated send and receive buffer pools. When the application requests a send buffer, MailRail pulls the buffer from the send queue. Once \text{MPI\_Test} confirms completion, MailRail recycles the buffer back to the send pool. Similarly, when the application posts a receive request, MailRail draws a buffer from the receive pool and releases the buffer when directed by the application. Environment variables configure the datagram size, pool sizes, and maximum outstanding receive requests.

\subsection{Flow Control}

Our MailRail implementation provides three flow control modes: unconstrained execution, rank-based flow control, and node-based flow control. In both node- and rank-based mechanisms, MailRail ties credits directly to available slots in the receive buffer pool. Because these credits reflect real receiver-side capacity, senders only transmit when space is available, preventing incoming messages from overwhelming receive buffers even in incast scenarios. An environment variable configures the credit-return threshold.

To prevent message communication from interfering with returning credits, MailRail duplicates the MPI communicator to create a dedicated credit communicator. MailRail returns credits using zero-payload messages, encoding the credit count directly in the message tag. This approach avoids buffer allocation and memory access during credit communication.

Both flow control mechanisms require a deferred send queue. When the application calls \text{mailrail\_send}, if the sender lacks sufficient credits for the destination process, MailRail places the send buffer into the deferred send queue and immediately returns control to the calling application. This design allows the application to continue executing without blocking on send credits. Whenever an application invokes \text{mailrail\_send} while messages remain in the deferred send queue, MailRail attempts to drain the queue by transmitting one message to each destination that now has available credits. MailRail then appends the new message to the queue, preserving FIFO ordering to prevent newer messages from bypassing older messages destined for the same target. 

The maximum number of outstanding send buffers (those in use by the application as well as those enqueued in the deferred send queue) is the number of slots in the send buffer pool. When all slots are in use, attempts to reserve buffers will result in backpressure until user-driven progress drains messages from the deferred send queue.

\subsubsection{Rank-Based Flow Control} 

In rank-based flow control, each sending rank begins with the same static number of credits. Credits are tied to the number of slots set aside for that sender in each receiver's receive buffer. The  number of credits per sender-receiver pair is: $$\text{Credits per pair} = 
\frac{\text{Total receive buffer size per receive rank}}{\text{Datagram size}\times \text{Total ranks}}$$

Rank-based flow control tracks credits symmetrically on both sender and receiver ranks. Senders maintain a credit count for each target destination, decrementing credits upon each transmission and sending only when credits are available. Conversely, receivers track consumed credits per sending rank, incrementing the count as they process and release incoming buffers. Once a sender’s consumed credits reach the configured threshold, the receiver returns them via the dedicated credit communicator.

MailRail also supports piggybacking credits in rank-based flow control. When transmitting a payload message, the sender checks for pending credit returns owed to that target rank. If credits exist, MailRail encodes the count directly into the message tag. Upon receiving any payload message, the receiver inspects the tag; if non-zero, it adds the tag value back to its send credits for that rank. An environment variable enables this piggybacking feature.

\subsection{Node-Based Flow Control}
In node-based flow control, MailRail tracks credits per node rather than per rank. The tracking mechanism mirrors rank-based flow control, using shared sender- and receiver-side credit counts per node. Before transmitting, a sender verifies that its local node has available credits, then decrements the shared node-level count. Upon receiving a message, the target process, on behalf of the receiving node, increments the credits consumed for the sending node. Once the consumed count reaches the return threshold, the receiving process transmits a credit-return message to the sending process, which then updates its node's sending credits. Because this design shifts tracking from rank pairs to node pairs, the credit allocation formula becomes:
$$\text{Credits per pair} = \frac{\text{Total receive buffer size per receive rank}}{\text{Datagram size}\times \text{Total nodes}}$$

There are three major advantages to tracking credits on a per-node basis rather than per-rank: reduced memory overhead, access to a large pool of shared credits, and a reduction in credit return message traffic. With rank-based flow control, the memory overhead for credit tracking is: $$2\times(\text{Ranks per node}\times\text{Node count})^2 \times \text{Bit width}$$
Because our node-based mechanism tracks credits in shared memory, updating adjacent credit entries on the same cache line can trigger false sharing and costly cache evictions. To prevent this performance degradation, we pad each credit entry to 128 bytes, isolating every credit entry onto its own cache line. Consequently, node-based flow control uses 
$$2\times(\text{Node count})^2 \times \text{128 Bytes}$$ to track credits. On 112-core processors, this design allows node-based flow control to use $425\times$ to $850\times$ less memory (depending on whether the bit width is 4 or 8 bytes) than rank-based flow control to track the same number of effective credits. Furthermore, node-based flow control enables receiver buffer pools smaller than one slot per rank without starving senders of credits or oversubscribing receiver memory.  

Under node-based flow control's shared credit model, active sending processes gain access to a larger pool of credits as neighboring processes on the same node finish their transmissions. Theoretically, this dynamic pooling is expected to shorten the latency tail relative to static rank-based flow control, without requiring receiver-side intervention.

Node-based flow control substantially reduces network overhead from credit-return messages compared to rank-based schemes. Because credit returns trigger when a threshold ratio of consumed-to-total credits is reached, aggregating credits at the node level increases the credit pool capacity by a factor of $k$, where $k$ is the number of ranks per node. This larger credit pool scales the credit-return threshold accordingly, offering a potential $k$-fold reduction in inter-node credit update messages, depending on traffic distribution. In addition, intra-node credit updates bypass the network entirely, resolving locally via low-overhead shared memory. Together, these mechanisms reduce fine-grained control traffic, reducing the risk of fabric congestion caused by frequent control packets.

\section{Experimentation}
\label{sec:evaluation}

\subsection{Hardware and Software}
Unless otherwise specified, all experiments are run on Dane, a supercomputer housed at Lawrence Livermore National Laboratory (LLNL). Dane contains 1,544 nodes, with each node having two 56-core, 2 GHz Xeon Platinum 8480+ (Sapphire Rapids) central processing units (CPUs) running the TOSS4 operating system. Nodes are connected via Cornelis Networks Omni-Path Architecture (OPA), using Performance Scaled Messaging 2 (PSM2) as the low-level communications library. The MPI library used is OpenMPI 4.1.2, and the compiler is GCC 10.3.1 using the default optimization levels.

Also unless otherwise noted, we use 4 KiB datagrams and 64 ranks per node in our experiments. We assign sending ranks 8 credits per receiving rank when using rank-based flow control. For node-based flow control, all ranks on sending nodes share 512 credits per receiving node. We conduct all rank-based flow control experiments using credit piggybacking. \trevor{not before submitting, but it would be good to document the effect this has} We set the credit return threshold to 2, meaning receiving processes return credits to senders after half of the senders' credits are consumed. This reduces the number of credit return messages sent while combating credit starvation.

\subsection{Adversarial}
We introduce an adversarial benchmark to study how incast communication affects performance. This benchmark represents an extreme example of incast communication, focusing entirely on communication and severely imbalanced workloads. The benchmark accepts four parameters: the number of victim processes, the datagram size, the number of messages sent per process, and the number of trials. It randomly selects victim processes from the process pool, after which each process randomly generates its message sending order. Processes aggressively receive communication before their first sent message, while waiting for available send buffers, and after each sent message. 

This benchmark performs no compute, calling \text{mailrail\_send} immediately after a successful buffer reservation. Victim processes do not send messages to themselves. Only the communication portion of the benchmark is timed, and we find goodput (the useful, application-level data transferred over a network, measured in bits per second) by dividing the total aggregate bits received by all victims by the overall runtime of the trial \cite{goodput}. All goodput numbers reported for node- and rank-based flow control schemes are based on this calculation method. 

Because incast communication degrades performance so severely, we add an option to capture sub-second execution statistics for granular insight into communication behavior. These metrics include elapsed time, sent message count, number of bytes sent, received message count, number of bytes received, completion percentages for sends and receives, the maximum time between \text{mailrail\_send} calls on the sender side, and the maximum time between \text{mailrail\_receive} calls on the receiver side. We print metrics per rank to avoid impacting runtime measurements. Note that because our benchmark collects these metrics at the application layer, it reports messages issued to \text{mailrail\_send}; under active flow control, these messages may still reside in MailRail's deferred send queue. This is why we use number of bytes received when reporting goodput for unconstrained execution.

\subsubsection{Weak Scaling}
To compare incast communication performance with and without flow control, we conduct two weak scaling experiments.  The first establishes a baseline for speedups between our methods of flow control and unconstrained execution by limiting the data sent per rank to a constant 12.8 MiB, which allows unconstrained execution on the adversarial benchmark to complete for up to 4096 ranks. The second experiment uses goodput to compare flow control and unconstrained execution at higher numbers of ranks; accordingly, we increase the amount of data sent to a constant 128 MiB per rank to maximize achieved goodput. We evaluate our adversarial benchmark using four victim ranks.
\paragraph{12.8 MiB} Incast communication over unconstrained execution degrades performance severely over PSM2, resulting in outright program crashes or continually decreasing throughput. Therefore, to establish a baseline comparison between unconstrained execution and our flow control methods, we limit the amount of data sent per rank to 12.8 MiB. This allows unconstrained execution to complete, enabling straightforward runtime comparisons. Runtimes reported are the average across five trials unless otherwise noted. No warm-up trial is used because runtimes remain largely consistent across trials.

Table~\ref{tab1} shows node- and rank-based flow control achieve a sizable speedup over unconstrained execution through an incast ratio of 1024:1. As workload doubles with the doubling of ranks, node-based flow control runtime also approximately doubles. Rank-based flow control nearly triples as workload doubles. In contrast, unconstrained execution begins at about 150$\times$ the runtime of either flow control method, and as workload doubles, unconstrained execution experiences a 5.7$\times$ to 8$\times$ increase in runtime. 

For node-based flow control, the resulting speedup over unconstrained execution is between 149$\times$ at an incast ratio of 128:1 and 4431$\times$ at an incast ratio of 1024:1. Rank-based flow control sees a more modest speedup over unconstrained execution, with speedups of between 156$\times$ at an incast ratio of 128:1 and 1902$\times$ at an incast ratio of 1024:1. 

Given their respective runtime growth rates, we expect node- and rank-based speedups over unconstrained execution to continue to increase at higher incast ratios. However, the runtime of unconstrained execution restricts this experiment to 4096 ranks; we project that one trial of 8196 ranks would take about 3 days and 16,384 ranks would take about 19 days to complete based on average runtime increases for unconstrained execution. 
\begin{table}[t]
\caption{Runtime Comparisons for 12.8 MiB}
\begin{center}
\begin{tabular}{|c|c|c|c|c|c|}
\hline
&\multicolumn{3}{c|}{\textbf{Average Runtimes (s)}} &\multicolumn{2}{c|}{\textbf{Speedup over No FC}}\\
\cline{2-6} 
 \textbf{Ranks}& \textbf{\textit{No FC}}& \textbf{\textit{Rank FC}}& \textbf{\textit{Node FC}}&
 \textbf{\textit{Rank}}& \textbf{\textit{Node}} \\
\hline
512& 152.4 & 0.979 & 1.02&155.7&148.9\\
\hline
1024& 895 & 2.55 & 2.3 &350.6&388.1\\
\hline
2048& 5080$^{\mathrm{a}}$ & 6.84 & 4.67 & 742.5 & 1086.8\\
\hline
4096& 40542$^{\mathrm{b}}$ & 21.3 & 9.15& 1902 & 4431\\
\hline
\multicolumn{6}{l}
{$^{\mathrm{a}}$Two trials completed in 3 hours (job time).}\\
\multicolumn{6}{l}
{$^{\mathrm{b}}$Two trials completed in 24 hours (job time).}
\end{tabular}
\label{tab1}
\end{center}
\end{table}
\paragraph{128 MiB}

To better measure the goodput achieved by node- and rank-based flow control, we increase the data sent per rank to 128 MiB. At these data loads, unconstrained execution at 512 ranks with an incast ratio of 128:1 takes an average overall runtime of 5463 seconds (about 1.5 hours), and 1024 ranks with an incast ratio of 256:1 finishes only one trial, with an overall runtime of 61,439 seconds (about 17 hours). Based on our previous observations, we could expect 2048 ranks to complete in over 4 days and 4096 to complete in over 3 weeks; since we evaluate flow control up to 16,384 ranks, running unconstrained execution at these scales is infeasible given allocation constraints, especially for a baseline known to behave poorly.

Consequently, to compare flow control to unconstrained execution at 128 MiB per rank, we use goodput to measure speedup and truncate execution for unconstrained runs only. Experiments conducted over PSM2 examining goodput over execution time for unconstrained execution show goodput tends to degrade further as time elapsed increases. We therefore cut off unconstrained execution after 600 seconds of sending and receiving data, then find the goodput using: $$\frac{\text{Aggregate Victim Bits Received}}{\text{Time (s)}}$$ This represents an upper bound on unconstrained execution goodput, especially since cutting off execution excludes tail latency from consideration. Flow-controlled execution is able to complete and hence includes tail latencies; we calculate flow control goodput using total aggregate bits received divided by overall completion time. 

This method is biased in favor of unconstrained execution, meaning flow control speedups reported are conservative estimates. This is confirmed by examining the actual runtime speedup of 512 ranks and 1024 ranks vs the truncated goodput speedups. The actual speedup for 512 ranks with an incast ratio of 128:1 is 628$\times$ for rank-based flow control and 655$\times$ for node-based flow control, while our truncation method only reports speedups of 287$\times$ and 299$\times$ respectively. For 1024 ranks with an incast ratio of 256:1, the actual speedup is 2776$\times$ for rank-based flow control and 3570$\times$ for node-based flow control, while the truncated method only reports speedups of 456$\times$ and 586$\times$ respectively.

\begin{figure}[t]
  \centerline{
  \includegraphics[width=\columnwidth]{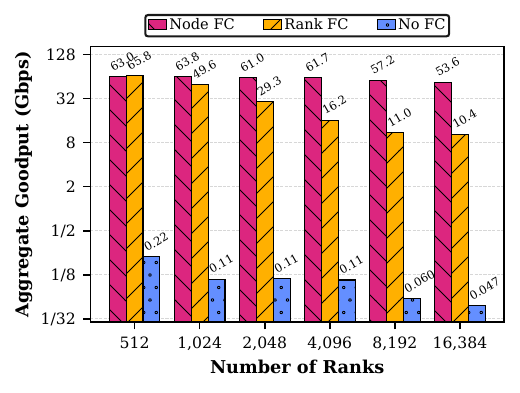}
  }
  \caption{Incast communication with 4 victim ranks. This histogram shows the goodput for incast ratios of between 128:1 and 4096:1 for node-based flow control, rank-based flow control, and unconstrained execution. Goodputs shown for unconstrained execution represent an optimistic estimate because execution is truncated and so lacks the latency tail. Note the y-axis grows exponentially.}
  \label{fig:adverse_weak}
\end{figure}

Of the mechanisms explored, node-based flow control performs best under incast communication, with application goodput only degrading by 14.9\% between an incast ratio of 128:1 at 512 ranks and a ratio of 4096:1 at 16,384 ranks (see Fig.~\ref{fig:adverse_weak}). Compared to our upper-bound goodput estimate for unconstrained execution, node-based flow control delivers speedups between 287$\times$ and 1132$\times$. 

Rank-based flow control performs 153$\times$ to 456$\times$ better than unconstrained execution. However, as rank size increases, aggregate goodput drops 84\% between incast ratios of 128:1 and 4096:1. Unconstrained execution experiences severe degradation under incast communication; unconstrained execution achieves an upper bound of 0.05\% to 0.01\% peak ingress throughput. This throughput failure is consistent with unregulated message injection saturating switch buffers, spreading congestion, and triggering hardware-level backpressure and slowly recovering source-rate limiting Our flow control methods regulate message injection at the process level, preventing incast traffic from exhausting fabric buffers.

\subsubsection{Incast Degree Sensitivity}
\begin{figure}[t]
  \centerline{
  \includegraphics[width=\columnwidth]{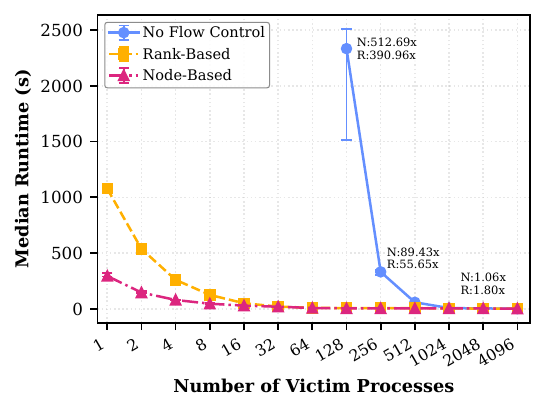}
  }
  \caption{Incast degree sensitivity across flow control mechanisms. This line graph shows a comparison of runtime performance relative to unconstrained baseline across varying communication patterns, from severe incast traffic (left, with an incast ratio of 4096:1) to uniform distribution (right, with a ratio of 1:1), with select speedup values.}
  \label{fig:victim_sweep}
\end{figure}

To evaluate flow control performance across the spectrum from severe incast to uniform traffic for pure communication, we use the adversarial benchmark to perform a victim sweep (see Fig.~\ref{fig:victim_sweep}). We hold the system scale constant at 4096 ranks and fix the aggregate data volume at 512 GiB. We define the victim sweep by varying the number of receiving (victim) ranks in power-of-two increments from 1 to 4096, transitioning from maximum incast severity (1 victim process) to a uniform distribution (4096 victim processes). Consequently, workload per victim rank doubles as the incast ratio increases.

Among the evaluated methods, node-based flow control demonstrates the greatest resilience to increasingly severe incast ratios while performing at least as well as unconstrained execution under uniform communication. Beginning at an incast ratio of 8:1, unconstrained execution exhibits approximately cubic runtime growth with respect to workload size: doubling the workload increases runtime by approximately 8$\times$. In contrast, node- and rank-based flow control maintain approximately constant runtimes until an incast ratio of 64:1. At that point, node- and rank-based flow control both begin to see linear runtime growth with respect to workload, with node-based flow control growing at a slower constant factor than rank-based.
At incast ratios of 4096:1 and 2048:1, node-based flow control achieves a 3.7$\times$ speedup over rank-based flow control.

Between 512 and 128 victim processes, node- and rank-based flow control exhibit significant performance speedups over unconstrained execution. At 512 victim processes (an incast ratio of 8:1), node- and rank-based flow control achieve a speedup of 15.8$\times$ and 10.8$\times$ over baseline respectively. This rises to 89$\times$ and  56$\times$ over baseline for 256 victims. At 128 victims (an incast ratio of 32:1) node- and rank-based flow control exhibit a 512$\times$ and 391$\times$ speedup, respectively. PSM2 timeout errors occur at 64 victims; therefore, we are unable to collect data on unconstrained execution beyond 128 victims.

Node-based, rank-based, and unconstrained execution exhibit similar performance for more uniform communication (victim counts 1024, 2048, and 4096). Even for these, however, node-based flow control demonstrates a 1.06$\times$ to 1.58$\times$ speedup compared to the unconstrained baseline. Rank-based flow control actually performs best in more uniform communication, with a speedup over baseline of between 1.47$\times$ to 1.66$\times$, with the uniform case being the highest speedup.

\subsection{Histogram}
Our histogram benchmark is set up similarly to YGM's histogram benchmark \cite{YGM}. The histogram table, of size $2^N$, is stored in distributed memory. Each process randomly generates integer entries to be incremented in the associated bin in the distributed table. To increase the effective bandwidth of the benchmark, we coalesce the 8-byte histogram entries on a per-receiver basis, using a user-specified datagram size. These messages are then sent via \text{mailrail\_send} to the process responsible for those entries \cite{histo}.

We use the histogram benchmark to evaluate the runtime overhead introduced by flow control under more ideal workloads (i.e., uniform communication paced through data coalescing), as well as to evaluate MailRail performance against YGM. We leave the default datagram size unchanged for YGM.  We use weak scaling for the histogram benchmark, with table size doubling as process size doubles and entries per rank (67,108,864, or 512 MiB) scaling with the number of processes. We run 50 trials per configuration. The initial trial of each batch is discarded as a warmup run. We calculate flow control overhead percentages by normalizing the flow control runtimes to the unconstrained baseline.

\begin{figure}[t]
  \centerline{
  \includegraphics[width=\columnwidth]{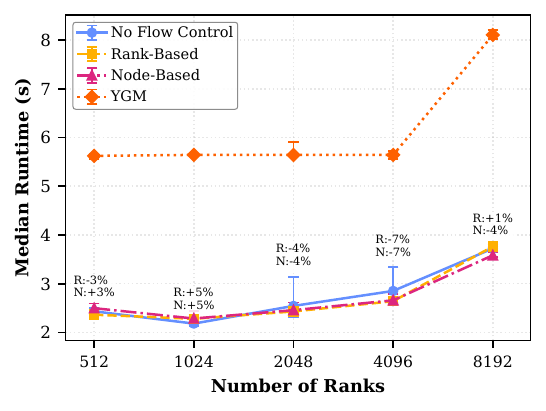}
mail  }
  \caption{Flow control runtime overhead comparison for the histogram benchmark across different ranks. Values represent median runtimes with min-max ranges across 50 trials.}
  \label{fig:hist_comp}
\end{figure}

 Both node- and rank-based flow control exhibit minimal overhead or even a slight improvement versus unconstrained execution. Fig.~\ref{fig:hist_comp} shows the runtime overhead of flow control across rank counts. Node-based flow control yields overheads ranging from $-6.7\%$ at 4096 ranks to $4.8\%$ at 1024 ranks, achieving a mean overhead of $-1.3\%$ overall. Similarly, rank-based flow control ranges from $-7.4\%$ overhead at 2048 ranks to $4.5\%$ at 1024 ranks, resulting in an average overhead of $-1.84\%$. On average, both node- and rank-based flow control match or modestly outperform unconstrained execution even with the histogram benchmark's more uniform workload. All of MailRail's configurations show a more than 2$\times$ speedup over YGM. This difference may reflect active-message and metadata overhead in YGM, although this study does not isolate those effects.

\subsection{Communication Trace Replay}

To test under real-world communications patterns and workloads, we introduce a replayer built on MailRail that reads and replays YGM traces. These are traces of recorded communication and omit computation. We replay traces on 8, 16, 32, 64, and 128 nodes, with 112 ranks per node.

\subsubsection{Reddit Sort Traces} 
\begin{figure}[t]
  \centerline{
  \includegraphics[width=\columnwidth]{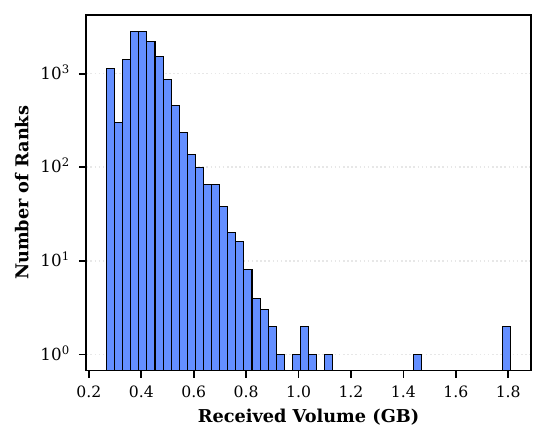}
  }
  \caption{Histogram showing how many ranks receive a given volume of data for Reddit sort traces at 14,336 ranks. Most processes receive between 0.3 and 0.6 GB of data, with outliers receiving up to 1.8 GB, indicating the communication pattern is irregular. }
  \label{fig:replay_sort}
\end{figure}

 To test performance of flow control on a dataset that exhibits irregular and imbalanced communication patterns (see Fig.~\ref{fig:replay_sort}), we replay Reddit comment sort traces that used strong scaling \cite{power-law,YGM}. The datagram size we use is 8 KiB. 
 
\begin{figure}[t]
  \centerline{
  \includegraphics[width=\columnwidth]{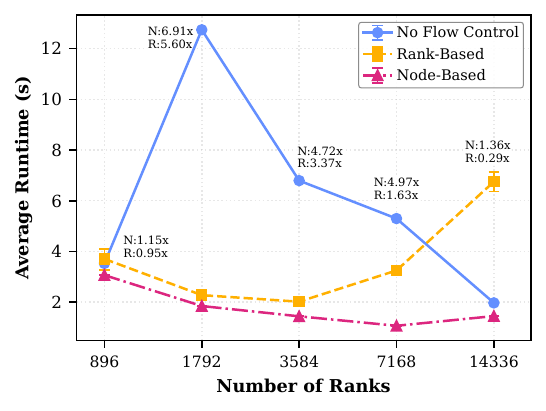}
  }
  \caption{Flow control performance comparison using replayed YGM Reddit sort traces. Annotations show speedup relative to unconstrained execution. At the largest rank scale, 8.6\% of total sender-receiver pairings have zero communication. This would prevent rank-based flow control from piggybacking credits for those rank pairs, which may partially explain its poor performance at the largest rank count.}
  \label{fig:replay_comp_sort}
\end{figure}

 Node- and rank-based flow control exhibit considerable speedups over unconstrained execution for all but 896 and 14,336 ranks (see Fig.~\ref{fig:replay_comp_sort}). While runtime is nearly the same on 896 ranks, for larger rank counts node-based flow control achieves a speedup between 1.36$\times$ and 6.91$\times$. Rank-based flow control has a speedup between 1.63$\times$ and 5.6$\times$ unconstrained execution for 1792 to 7168 ranks; at 14,336 ranks, rank-based flow control performs nearly 4$\times$ worse than unconstrained execution. 
 
 Potential reasons for rank-based flow control's poor performance at 14,336 ranks include receiver buffer underutilization and control message traffic. At this scale, Reddit sort traces have 17.6 million sender-receiver pairings with zero communication, representing 8.6\% of all pairs. Because rank-based flow control allocates credits to every sender-receiver pair in advance, pairs with no communication retain credits tied to receive-buffer capacity that active rank pairs cannot use. This can lead to unnecessary credit starvation in active senders that would benefit from access to these unused receive buffers. Node-based flow control naturally shares credits across processes and therefore avoids this limitation.
 
 Another factor contributing to node-based flow control’s superior performance over rank-based flow control may be its credit-return strategy; while rank-based flow control averages between 17.8 million and 26.4 million credit return messages per trial at 14,336 ranks, node-based flow control averages between 116,922 and 435,055, reducing network congestion caused by control messages.


\subsubsection{Triangle Counting Traces}

\begin{figure}[t]
  \centerline{
  \includegraphics[width=\columnwidth]{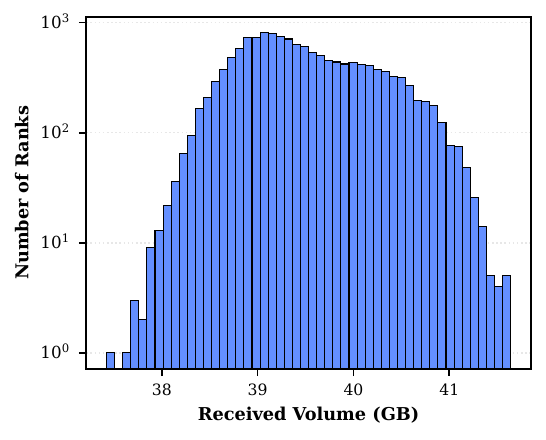}
  }
  \caption{Histogram showing the how many ranks received a given volume of data for triangle counting traces at 14,336 ranks. Most ranks received between 38 GB and 41.5 GB of data, indicating that overall triangle counting communication is relatively uniform. }
  \label{fig:replay_tri}
\end{figure}

Triangle counting is a common benchmark to test irregular inter-node communication in HPC settings. We replay triangle counting traces created using R-MAT graphs \cite{mit, RMAT}. This triangle counting uses weak scaling, beginning with $2^{25}$ vertices and about $2^{25}\times16$ edges for 8 nodes. For each doubling of compute nodes, the number of vertices scales by a power of 2, up to $2^{31}$ vertices. Aggregate data replayed ranges from 3.6 TB on 896 ranks to 112.7 TB on 14,336 ranks. At 128 nodes (14,336 ranks), each rank receives a total of between 38 and 41.5 GB of messages, indicating that overall triangle counting communication is relatively uniform at this rank count (see Fig.~\ref{fig:replay_tri}).

Node- and rank-based flow control achieve a small speedup over unconstrained execution for lower rank counts (between 1.07$\times$ and 1.53$\times$). However, at 14,336 ranks, node- and rank-based flow control exhibit a speedup of about 10$\times$ over unconstrained execution, suggesting that flow-control benefits can extend beyond extreme incast workloads.

\begin{figure}[t]
  \centerline{
  \includegraphics[width=\columnwidth]{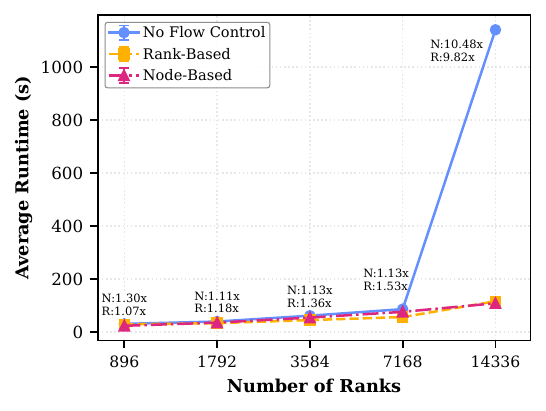}
  }
  \caption{Flow control performance comparison using replayed YGM traces. Annotations show speedup relative to unconstrained execution.}
  \label{fig:replay_comp}
\end{figure}

\subsection{OPX Comparison}
We also evaluate our adversarial benchmark using only rank-based flow control on Lynx. Lynx is an experimental supercomputer running on the Cornelis CN5000 Omni-Path fabric using Omni-Path Express (OPX). OPX uses receiver-driven flow control, and CN5000 Omni-Path handles incast communication by pacing senders using telemetry to modulate injection rates \cite{cornelis_whitepaper}. We run our benchmark for 8, 16, 32, 64, 128, 256, 512, and 768 nodes, using 64 ranks per node.

\begin{figure}[t]
  \centerline{
  \includegraphics[width=\columnwidth]{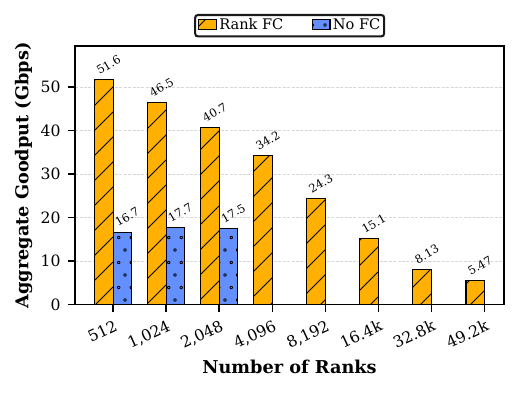}
  }
  \caption{Rank-based flow control vs unconstrained execution on Lynx over OPX, which implements receiver-driven flow control. This was run on 8, 16, 32, 64, 128, 256, 512, and 768 nodes, with 64 ranks per node. Unconstrained execution experiences OOM errors beginning at 1024 ranks. Rank-based flow control is able to complete execution at the maximum rank count tested.}
  \label{fig:opx_comp}
\end{figure}

At the time of our experiment, unconstrained execution on Lynx experienced OOM errors beginning at 1024 ranks, with some but not all trials able to complete at 1024 and 2048 ranks (see Fig.~\ref{fig:opx_comp}). Rank-based flow control achieves 3$\times$ speedup over unconstrained execution at an incast ratio of 128:1 and 2.3$\times$ at a ratio of 512:1. Rank-based flow control at an incast ratio of 4096:1 has approximately the same goodput as unconstrained execution at an incast ratio of 512:1. Rank-based flow control is able to complete execution on 49,152 ranks, with an incast ratio of 12,288:1, though trials take about 2.75 hours to complete.

\section{Summary and Conclusion}
\label{sec:conclusion}

Node-based flow control delivers substantial gains on the evaluated adversarial incast workloads and incurs minimal overhead, or provides modest improvements, on the evaluated uniform and trace-replay workloads.

Node-based flow control delivers substantial performance gains on the evaluated adversarial incast workloads and incurs minimal overhead, or provides modest improvements, on the evaluated uniform and real-world communication patterns, as demonstrated through our evaluation of MailRail on synthetic benchmarks and YGM application traces. By avoiding the static credit-allocation constraints of rank-based flow control and regulating message injection relative to unconstrained execution, node-based flow control delivers up to a $1132\times$ speedup over unconstrained execution at an adversarial incast ratio of 4096:1, sustains goodput even under extreme 4096:1 incast scenarios, and exhibits linear runtime scaling as victim workloads increase. Ultimately, given its low overhead and consistently high goodput, node-based flow control provides a scalable approach to mitigating incast communication.

\section*{Acknowledgment}
This work was performed under the auspices of the U.S. Department of Energy by Lawrence Livermore National Laboratory under
Contract DE-AC52-07NA27344. 
This material is based upon work supported by LLNL LDRD 
24-ERD-047; LLNL-CONF-2022958.

The authors would like to thank Nate Hanford for his assistance running jobs on Lynx, as well as for generously sharing his ample knowledge.
\bibliographystyle{IEEEtran}
\bibliography{references}
\end{document}